\documentstyle[12pt]{article}
\newcommand{\be}{\begin{equation}} \newcommand{\ee}{\end{equation}}
\newcommand{\bea}{\begin{eqnarray}}\newcommand{\eea}{\end{eqnarray}}

\begin{document}
\begin{titlepage}
\title { A Heavy Glueball in a Bag Model at
Finite Temperature}
\vskip 3.0 cm
\author{ {\bf Lina Paria}$^{1}$, {\bf M. G. Mustafa}$^{2}$ and
{ \bf Afsar Abbas}$^{1}$  \\
 $ ^{1}$Institute of Physics, Sachivalaya Marg, \\
Bhubaneswar-751005, India. \\
$ ^{2} $Variable Energy Cyclotron Centre, 1/AF Bidhan Nagar \\
Calcutta - 700 064. India.}
\footnotetext[1]{e-mail: \ lina/afsar@iopb.ernet.in}
\footnotetext[2]{e-mail: \ mustafa@veccal.ernet.in}
\maketitle
\thispagestyle{empty}
\begin{abstract}
We obtain a heavy glueball (much heavier than the ones studied by
others which usually are in the range of 1-2 GeV) in a bag model
calculation with exact discrete single particle states of gluons
 at finite temperature. This heavy glueball, within the cosmological
context, is what Abbas has recently predicted (hep-ph/9504430).

\end{abstract}
\end{titlepage}
\eject

\eject

\newpage

Since the establishment of the quark model [1] and Quantum
Chromodynamics QCD [2], the nonobservability of quarks and
gluons has necessitated the concept of confinement.
QCD is non-perturbative at large distances.
The confinement itself can not be treated perturbatively and
is yet to be confirmed analytically.
However lattice calculations[3] do indicate confinement. Based on
confinement and due to the non-abelian nature of QCD, bound states
of gluons, the so called glueballs, should exist.
Much work has been done both theoretically [4,5] and experimentally [6]
on the glueballs which suggests  that the lowest state glueball
must be light ( $1 - 2$ GeV ). Recently Abbas [7] has suggested that
if we take the two distinct QCD phase transitions [8] in the early
universe, wherein the gluon transition temperature is $ T_g \sim 400 $ MeV
and the quark transition temperature is $ T_q \sim 250$ MeV, then as the
universe cool the gluon condensates to a weakly interactng
massive glueball ( $ \ge 45 $ GeV). These then decouple from the rest
of the universe. Hence they become a natural candidate for the Dark-Matter
(DM). Here we wish to understand this prediction of the existence of a
heavy glueball in another framework discussed below.

A study of the glueballs was performed [9 -14] in the framework of the
MIT Bag model. Both the lattice Monte - Carlo [15] and the
Bag model [9 -14] approaches predict a large number of glueballs between
$ 1 - 2 $ GeV. In these bag studies, the bag constant $ B^{1 \over 4}$ was
taken maximum  upto $ 230 $ MeV. The finite size corrections
to the bag containing
only gluons were calculated by Jennings and Bhaduri and for quarks by
Bhaduri et al. [16]. They calculated the effective smooth single particle
density of states for gluons and quarks to include the finite size
effects. They used an approximation. The discrete sum over single particle
states in a finite bag was approximated by an integral with single
particle density of states.
These calculations are valid within the domain $ RT \ge 1$,
 $ R $ is the radius of the bag and $ T $ is
the temperature of the system. Later Dey et al. [17] calculated the
free energy $ (F) $ and the other thermodynamic quantities for the
quark - gluon system by computing the discrete allowed states and
taking appropriate weighted sums. In this discrete sum method, they
have discussed the first order phase transition from hadrons to QGP.
Later Ansari et al.[18] studied the quark - gluon system within a
spherical bag and as well as in a deformed bag of spheroidal shape.
They constructed a grand canonical partition
function in terms of quark-gluon single particle states
inside a bag.
 They found [18] a temperature range  $T_s \ < \ T \ < \ T_c$
in the $ \mu - T $ plane in which a superheated (supercooled) metastable
state exists. Beyond $ T_c $ there is no bag solution, indicating
a first order phase transition from hadrons to QGP.

In this paper we plan to study the thermodynamics of a bag containing
 only gluons using the discrete allowed states in such a bag.
These discrete single particle states have been obtained by solving the
equation of motion with linearised boundary conditions in a MIT bag[19].
The solutions for the gluonic bag $ R(T) $ is obtained from the
extrema of the total free energy $ (F_T) $. The two extrema in $ F_T $
correspond to a light glueball and a heavy glueball.
For a particular bag pressure constant $ B $, there is a transition
temperature (say $ T_s $) below which only the low mass excitation
 exists. For $ T_s \le T < T_c $  the two extrema case
arises. There is a critical temperature $ T_c $ above which no bag
solution $ R(T) $ exists indicating the deconfinement of glueballs.

In this model we consider that a system of gluons is placed in a heat bath
with which it can exchange the energy and the particle number through
which it achieves thermal equilibrium. We study the thermodynamic
properties of a thermalized system.

The partition function  for the gluonic system is given by,
\be
ln Z \ = \ - \ 8 \ {\sum_i} \ {g_i} \ ln { \Big ( } 1 \ - \ e^{-{{\epsilon_i}
\over T}} {\Big )}
\ee

\noindent
where $ g_i $ is the spin degeneracy factor for the
$ i^{th} $ single particle state with energy $ \epsilon_i $. The degeneracy
factor $ 8 $ arises due to the $ SU(3) $ color - group in which gluons
correspond to adjoint representation.

The energy and the free energy of the gluonic system are

\bea
 E \  & = &  \ T^2 \ {\Big(} {\partial lnZ \over
{\partial T}} {\Big)}  \nonumber \\
\ & = & \ 8  \ {\sum_i} \ {{g_i} \ {\epsilon_i}
\over (\ e^{{\epsilon_i} \over T} - 1)}
\eea

\noindent

\bea
F \ &  = &  \ - T \ ln Z \nonumber \\
\  & = & \ 8 T \ {\sum_i}  \ {g_i}
\ ln {\Big (} \ 1  \ -  \ e^{- {\epsilon_i \over T}} {\Big)}
\eea
\noindent
If we include the zero point energy of the bag [19]
$ BV \ + \ {C \over R} $
to Eq.(2) and (3), then the total energy and the free energy of the bag
becomes

\be
 E_{T}  \ =  \ E  \ +  \ BV  \ + \ {C \over R}
\ee

\noindent

\be
 F_{T}  \ = \ - \ T \ lnZ  \ +  \ BV  \ + \ {C \over R}
\ee
\noindent
where the values of $ C  \sim  0.36 $ is taken from ref.[20]
and $ B $ is the conventional bag pressure constant and $ V$ is the
volume of the system.

The pressure generated by gluons inside the bag
\bea
P_{gluons} \  & = &  \ - \left ({\partial F_{T} \over
{\partial V}}\right)_T  \nonumber \\
 \ & = & \ {1 \over 3 V } ( E \ + \ { C \over R } )
\eea

\noindent

Where $ V \ = \ {4 \over 3} \ \pi \ R^3 $ \\

As we know, for a stable configuration[19], the total pressure on the
surface of the bag vanishes. \\
$ P \ = \ P_{gluon} \ - \ B \ = \ 0 $ \\
Hence,
\be
 P_{gluons} \ = \ B
\ee

\noindent

So at the extremum of $F_{T} $ where the total pressure vanishes,
the energy is given by,
\be
 E_{T} \ =  \ 4BV
\ee

\noindent

The number of gluons, $ N_{G} $ of the glueball at a given $ T $ can be
obtained as,

\be
N_{G} \ = \ 8 \ \sum_{i} { g_i} {f_i}
\ee
\noindent

where $ f_i \ = \ {1 \over \  {\Big(} \ e^{\epsilon_{i} \over T}
\ - \ 1{\Big)}} $ , the B - E distribution function.

Now treating the gluonic bag(glueball) at high $ T $ like a many-body
system, its stability features are studied by plotting $ F_{T} $ as a
function of bag radius $ R $, since the physical behaviour of the
system at a given $T$ is governed by the properties of its free energy
$F_{T} $. In Fig.1. a plot of total free energy($F_{T}$) vs. $R$
is displayed for $ B^{1 \over 4} \ = \ 250 $ MeV.

 We see that there is a transition temperature $ T_s $ below
which $ F_T $ has only a single minimum with a finite $ R $ value. But
at the temperature $ T = T_s = 218.5 \ MeV $,
 $ F_T $ has two extrema, one with smaller $ R $ and the other one is with
larger $ R $. Now as the temperature increases, both the solutions approach
each other and at the critical temperature $ T = T_c = \ 256.9 \
 MeV \ > T_s $,
both solutions meet at one point and beyond $T_c $ there is no
extrema in $ F_T $ implying no bag solution and glueball does not exist.
In Fig.2. and Fig.3. we plot the radius of bag $ (R) $ and mass ($M$)
corresponding to the extrema
of $ F_T $ as a function of the temperature $ T $. We see from Fig.2.
\&  3. that the radius corresponding to minima in $F_T$
remains almost same $(0.335 fm) $ upto $T_s$ ($218.5$ MeV)
giving a stable state with  very low mass ($0.32$ GeV).

Since physically a meaningfull maximum appears in $F_{T} $ at $T_s$ with
$ R \ = \ 4.08 \ fm $ (Fig.2), indicating a heavy metastable glueball
with very high mass ($\sim 578 \ GeV $) (Fig.3). As the stable state
has very low mass compared to the heavy metastable state at $ T \ = \ T_s $,
there is a sharp increase in the Fig.3. Now in the region $ T_s \ < \ T \
< \ T_c $ the solution with lower $ R$, corresponding to minimum in $F_T$
(Fig.1.), expands very slowly (in Fig.2.) with slow increase of mass
(Fig.3.) whereas the solution with larger $R$, corresponding to maximum
in $F_T$ (Fig.1.), contracts very fast (Fig.2) with the rapid decrease
of mass(Fig.3). Since at $T \ = \ T_c $, both solutions converge at
one point with $ R \ = \ 0.422 \ fm $ and mass $ \sim 0.64\ GeV $.

Note that for $ B^{1 \over 4}=400 \ MeV $, $ T_s $ becomes
$ \sim \ 350 \ MeV $ with a heavy metastable glueball
having mass $ \sim 670 \ GeV $ whereas
that of the light glueball is $ \sim 0.5 \ GeV $. The value of $ T_c $ turns
out to be $ \sim 410 \ MeV$ with  $ R \sim 0.25 \ fm $ and mass
$\sim  1 \ GeV $.

The heavier mass state tends to be highly collective, e.g. for
$ T=220 \ MeV $ with $ B^{1 \over 4} = 250 \ MeV $,
 the ratio of the gluon number (gluon number is calculated using
$ N_G$) in the heavy  mass glueball
state to that of the low mass glueball state is $ 10605 $.
Though the collectivity does go down with the increase in temperature,
e.g. at $ T=230 \ MeV $, the ratio comes down to $ 385 $ and at
$ T=240 \ MeV $, the ratio is $ 62 $. One may speculate that
this large collectivity of the heavy glueball may lead to a greater
stability of the same.

The most interesting result here is the existence of the very
massive glueball. This is exactly what was predicted by
Abbas[7] recently in the context of his cosmological arguments. We feel
that our calculations here should have much relevance to the early
universe scenario as well as in QGP. This calls for further study.

\vfill
\newpage
\begin{figure}

\noindent {\centerline{\bf FIGURE CAPTIONS}} \\
\vskip 1.0 true cm

\caption{ The variation of the total free energy $ (F_{T}) $ as a function
of the radius ($ R $) of the gluonic system for different
temperatures with $ B^{1 \over 4} = 250 \ MeV $ is shown. Here
$ T_s \ = \ 218.5 \ MeV $ and $  T_c \ = \ 256.9 \ meV $.}

\caption{ The variation of the radius $(R) $ of the system
with temperature ($T$) is shown. Here $ B^{1\over 4} = 250$ MeV,
$ T_s = 218.5$ MeV and $ T_c = 256.9$ MeV .}

\caption{The mass $ (M) $ variation of the gluonic system with
temperature $ (T) $ is displayed.}
\end{figure}
{}~
\vfill

\end{document}